\documentclass[english]{article}

\usepackage[latin9]{inputenc}
\usepackage{geometry}
\usepackage{babel}
\usepackage{float}
\usepackage{units}
\usepackage{amsmath}
\usepackage{amssymb}
\usepackage{graphicx}
\usepackage[unicode=true]
 {hyperref}

\makeatletter

\floatstyle{ruled}
\newfloat{algorithm}{tbp}{loa}
\providecommand{\algorithmname}{Algorithm}
\floatname{algorithm}{\protect\algorithmname}

\newcommand{\lyxaddress}[1]{
	\par {\raggedright #1
	\vspace{1.4em}
	\noindent\par}
}

\usepackage{algorithm}
\usepackage{algpseudocode} 
\usepackage{ae,aecompl}

\makeatother

\begin{document}
\title{Statistical spatial analysis for cryo-electron tomography}
\author{Antonio Martinez-Sanchez\textsuperscript{1,2,3,4,5{*}}, Wolfgang
Baumeister\textsuperscript{5} and Vladan Lu\v{c}i\'{c}\textsuperscript{5,{*}}}
\maketitle

\lyxaddress{\textsuperscript{1}Department of Computer Sciences, Faculty of Sciences
- Campus Llamaquique, University of Oviedo, 33007 Oviedo, Spain\\
\textsuperscript{2}Health Research Institute of Asturias (ISPA),
Avenida Hospital Universitario s/n, 33011 Oviedo, Spain\\
\textsuperscript{3}Institute of Neuropathology, University Medical
Center Göttingen, Göttingen, Germany\\
\textsuperscript{4}Cluster of Excellence \textquotedblleft Multiscale
Bioimaging: from Molecular Machines to Networks of Excitable Cells\textquotedblright{}
(MBExC), University of Göttingen, Göttingen, Germany\\
\textsuperscript{5}Department of Molecular Structural Biology, Max
Planck Institute for Biochemistry, Am Klopferespitz 18, 82152 Martinsried,
Germany\\
\textsuperscript{{*}}E-mail: martinezsantonio@uniovi.es, vladan@biochem.mpg.de}
\begin{abstract}
Cryo-electron tomography (cryo-ET) is uniquely suited to precisely
localize macromolecular complexes \textit{in situ}, that is in a close-to-native
state within their cellular compartments, in three-dimensions at high
resolution. Point pattern analysis (PPA) allows quantitative characterization
of the spatial organization of particles. However, current implementations
of PPA functions are not suitable for applications to cryo-ET data
because they do not consider the real, typically irregular 3D shape
of cellular compartments and molecular complexes. Here, we designed
and implemented first and the second-order, uni- and bivariate PPA
functions in a Python package for statistical spatial analysis of
particles located in three dimensional regions of arbitrary shape,
such as those encountered in cellular cryo-ET imaging (PyOrg).

To validate the implemented functions, we applied them to specially
designed synthetic datasets. This allowed us to find the algorithmic
solutions that provide the best accuracy and computational performance,
and to evaluate the precision of the implemented functions. Applications
to experimental data showed that despite the higher computational
demand, the use of the second-order functions is advantageous to the
first-order ones, because they allow characterization of the particle
organization and statistical inference over a range of distance scales,
as well as the comparative analysis between experimental groups comprising
multiple tomograms.

Altogether, PyOrg is a versatile, precise, and efficient open-source
software for reliable quantitative characterization of macromolecular
organization within cellular compartments imaged \textit{in situ}
by cryo-ET, as well as to other 3D imaging systems where real-size
particles are located within regions possessing complex geometry.
\end{abstract}

\section{Introduction}

The last decades of research in cell biology have revealed that cellular
processes are performed by groups of interacting macromolecules in
a crowded environment. This is in contrast to earlier models where
macromolecules were considered to exist as isolated objects floating
randomly in the cytoplasm. Therefore the analysis of their organization
within their native cellular compartments can provide quantitative
information that can be used to describe the mechanisms underlying
macromolecular interactions. This information has paramount importance
to gain a deeper understanding of various cellular interactions \cite{Mitrea2016,Rosenzweig2017}.

Cryo-Electron Tomography (cryo-ET) is a unique imaging technique capable
of producing 3D views of large portions of cells at a resolution that
is sufficiently high to localize and identify macromolecular complexes
\cite{Lucic2013}. In cryo-ET, biological samples are vitrified to
preserve their natural molecular organization and are imaged by electron
microscopy in the fully-hydrated vitrified state, thus enabling the
study of cells in a close-to-native state at high resolution \cite{Taylor:74,Dubochet1988}.

Point patterns analysis (PPA) is a branch of statistics devoted to
describing quantitatively point patterns in space. PPA has been used
extensively in experimental fields such as ecology \cite{Wiegand2004},
social sciences \cite{Marcon2003,Duranton2005}, and more recently
in biology \cite{Plowman2005,Owen2010}. Among PPA methods, first-order
functions describe clustering of points and typically determine a
characteristic distance scale, whereas second-order functions determine
spatial correlations between points at a range of distances. Ripley's
functions $K$ (or its linearization, $L$) and $O$ \cite{Wiegand2004}
are the most used tools for second-order PPA. They have been used
for analyzing experimental biological data obtained from light and
electron microscopy \cite{Owen2010,Andrey2010,JafariMamaghani2010}.
Current implementations of the PPA methods were either developed for
2D spaces \cite{Wong2005}, or their application in 3D is limited
to sphere-shaped structures within cubic volumes \cite{Andrey2010,Hansson2013}.
However, these approaches have two limitations, rendering their application
impractical for cellular cryo-ET data: (i) cellular compartments have
complex geometry so, there are no analytical solutions for border
compensation, unlike for the most simple geometries \cite{Goreaud1999,Hansson2013},
and (ii) proteins have specific shapes that can not be properly represented
by points or spheres. Recently Voronoi tessellation has been used
to describe the particle organization in super-resolution optical
microscopy \cite{Andronov2018}, but contrarily to Ripley's functions
it does not allow comparison between datasets having different density
of points.

The authors have already applied PPA functions to successfully solve
some long-standing biological questions. Firstly, in \cite{Rosenzweig2017},
we applied a modification of Ripley's function $O$ to prove that
Rubisco complex has a liquid-like organization within the pyrenoid
organelle, thus discarding crystalline models proposed previously.
This finding was also remarkable as cryo-ET data was used to successfully
analyze a phase-separated compartment \textit{in situ} at nanometric
resolution for the first time. We also showed that the Arp2/3 complex
within actin waves modifies its clustered organization depending on
the wave-phase \cite{Jasnin2019}. Recently, we demonstrated that
Ripley's functions can be used to characterize nano-domains formed
by synaptic membrane-bound complexes \cite{Martinez2021}.

Here we present an implementation of the first and second order PPA
functions where both particles and cellular compartments can be of
arbitrary shape. Consequently, these functions are applicable to cryo-ET
images of cellular samples, that is they are suitable for the spatial
distribution analysis of individual proteins or macromolecular complexes
localized in any kind of cellular compartment (cytoplasm, organelle
lumen, membrane, etc). These numerical calculations are required to
precisely calculate the computationally intensive second order PPA
functions. We also present a parallel implementation suitable for
processing cryo-ET datasets that takes advantage of modern multiprocessor
architectures \cite{Fernandez2012}. In addition to univariate PPA
functions, we also implemented the bivariate versions of PPA functions,
which enable the colocalization analysis between different proteins
and macromolecules.

We present several applications to synthetic datasets in order to
compare different numerical methods and find the most appropriate
ones. Finally, we also applied the methods we implemented on real
Cryo-ET datasets to validate their real-case usability and to justify
the necessity for an accurate implementation of second-order PPA function.

\section{Approach}

In this section, we show the implementation details relevant for the
first and second order PPA functions, for univariate and bivariate
cases.

\subsection{Design}

\subsubsection{Monovariate first-order analysis}

Nearest neighbor function $G$, spatial contact distribution function
$F$, and their combination $J$ belong to the first order PPA functions.
Location of proteins or macromolecular complexes of interest (particles)
in a tomogram (3D image) is defined by their spatial coordinates $\mathbf{X}=\left\{ \mathbf{x}_{i}\in\mathcal{V}\:\forall i=(1,\ldots,n)\right\} $,
where $n$ is the number of particles and $\mathcal{V}\in\mathbb{R^{\mathrm{\mathit{d}}}}$
is the Volume of Interest (VOI), that is a subspace of the tomogram
where the particles are located. A VOI typically represents an organelle
or a distinct cellular compartment.

Function $G$ is defined as:
\begin{equation}
G(r)=\int_{0}^{r}p_{G}(r)dr
\end{equation}
where $p_{G}(r)$ is the probability distribution function of nearest
neighbor distances $r$ of all particles in $\mathbf{X}$.

Function $F$ is similarly defined, except that it requires $p_{F}(r)$,
the probability distribution function of nearest neighbor distances
among points in sets $\mathbf{X}$ and $\mathbf{X^{CSR}\in}\mathcal{V}$,
where $\mathbf{X^{CSR}}$ is a set of coordinates distributed according
to the Complete Spatial Randomness (CSR) model in $\mathcal{V}$ \cite{Wiegand2004,Andrey2010}:
\begin{equation}
F(r)=\int_{0}^{r}p_{F}(r)dr
\end{equation}

The calculation of functions $G$ and $F$ are not computationally
demanding even though function $F$ requires the generation of set
$\mathbf{X^{CSR}}$. $G$ is better suited to characterize short scale
properties of particle clusters, while $F$ characterizes the space
devoid of particles and large scale organization of particle clusters.
In order to combine their advantages, function J was defined as:
\begin{equation}
J(r)=\frac{1-G(r)}{1-F(r)}
\end{equation}

However, the practical applications of function $J$ are limited because
it is not defined for $F(r)=1$ and numerical problems appear when
$F(r)\rightarrow1$.

\subsubsection{Monovariate second-order analysis}

\begin{figure}
\begin{centering}
\includegraphics[width=1\textwidth]{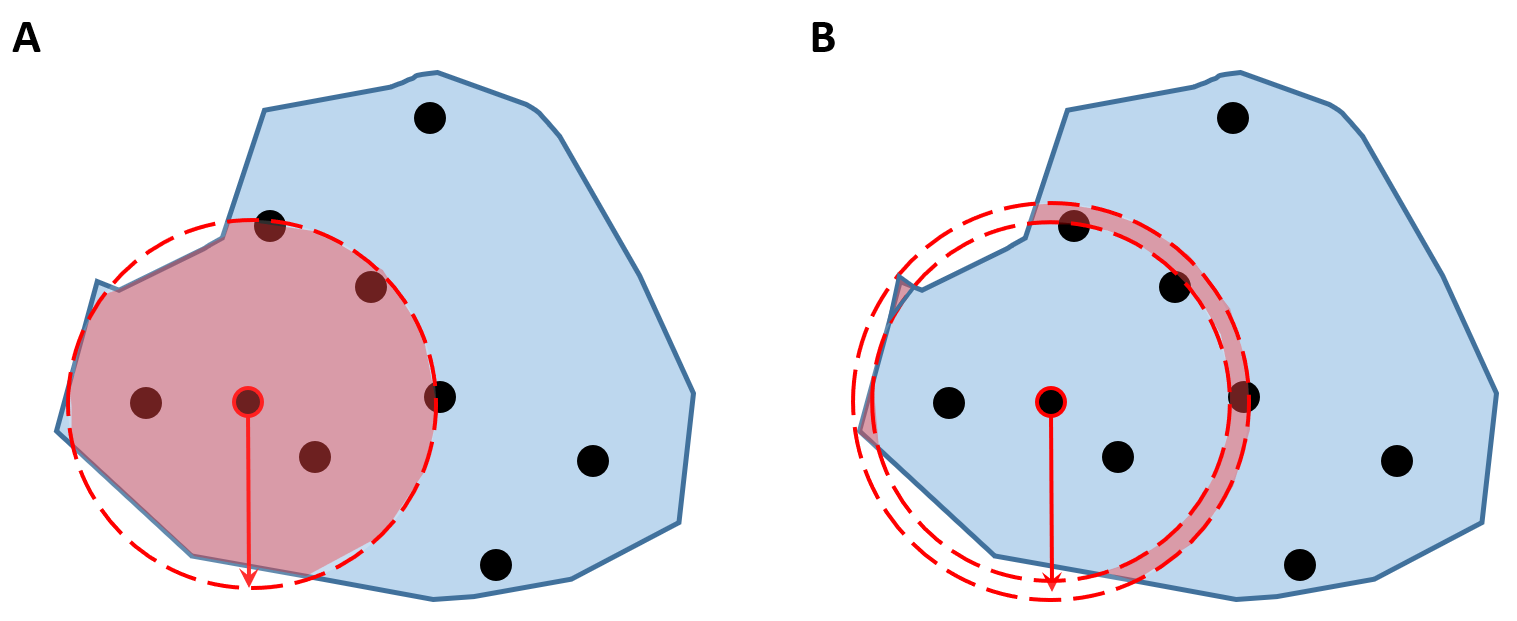}
\par\end{centering}
\caption{\label{fig:2d-schemes}Particle neighborhood and shell for Ripley's
functions. (A) Edge corrected particle neighborhood $\mathcal{S}_{L}$
used for Ripley's $L$ is formed as the intersection of the particle
spherical neighborhood (shown as the red circle) and VOI (blue), and
is indicated by light red areas. (B) Edge corrected particle shell
$\mathcal{S}_{O}$ used for Ripley's $O$ is formed as the intersection
of the particle spherical shell (shown as the red ring) and VOI (blue),
and is indicated by light red areas. In both cases particles are represented
by black points. Shown in 2D for clarity.}
\end{figure}

Ripley's functions $K$, $L$ and $O$ are used for second-order analysis.
Because cellular compartments (VOIs) typically have an irregular shape
and many particles are found close to the compartment borders, it
is not possible to determine edge-corrections for Ripley's functions
analytically \cite{Goreaud1999,Wiegand2004}. Therefore, we proceeded
to implement Ripley's function where edge-corrections are performed
numerically, for each particle separately. Specifically, function
$L$ is obtained by the linearization of Ripley's $K$ function \cite{Wiegand2004},
which facilitates the interpretation. For a set of particles located
at positions $\mathrm{\mathbf{x}}_{i}$ within a 3D VOI of arbitrary
shape $\mathcal{V}$ , Ripley's functions $K$ and $L$ are defined
as follows:
\begin{equation}
\begin{array}{cc}
K(r)=\frac{4\pi r^{3}}{3}\cdot\frac{\sum_{i=0}^{n}C(\mathbf{x},\mathcal{S}_{L}(\mathbf{x}_{i},r))}{\lambda\cdot\sum_{i=0}^{n}\mathcal{\mathit{V}}(\mathcal{S}_{L}(\mathbf{x}_{i},r))} & \;L(r)=\sqrt[3]{\frac{3K(r)}{4\pi}}-r\end{array}\label{eq:ripleys_L}
\end{equation}
$C(\mathbf{x},\mathcal{S}_{L}(\mathbf{x}_{i},r))$ is the number of
particles located in the neighborhood $\mathcal{S}_{L}(\mathbf{x}_{i},r)$
and $\mathcal{\mathit{V}}(\mathcal{S}_{L}(\mathbf{x}_{i},r))$ is
the volume of this neighborhood. Neighborhood $\mathcal{S}_{L}(\mathbf{x}_{i},r)$
is defined as the edge-corrected neighborhood of the particle located
at $\mathrm{\mathbf{x}}_{i}$, which is obtained by the intersection
of the VOI and the spherical neighborhood of the particle at radius
$r$: 
\begin{equation}
\mathcal{S}_{L}(\mathbf{x}_{i},r)=\left\{ \forall\mathbf{x}\in\mathbf{\mathcal{V}}|d(\mathbf{x},\mathbf{x}_{i})\leq r\right\} 
\end{equation}

\noindent where $d$ is a distance metric (Fig. \ref{fig:2d-schemes}A).
In an unbounded space $\mathcal{V\mathrm{=\mathbb{R}^{3}}}$, or for
particles located far from boundaries, $\mathcal{S}_{L}(\mathbf{x}_{i},r)$
is simply a sphere centered at $\mathbf{x}_{i}$ with radius $r$.

Ripley's function $O$ is defined as:
\begin{equation}
O(r)=\frac{\sum_{i=0}^{n}C(\mathbf{x},\mathcal{S}_{O}(\mathbf{x}_{i},r,\triangle r))}{\sum_{i=0}^{n}\mathcal{\mathit{V}}(\mathcal{S}_{O}(\mathbf{x}_{i},r,\triangle r))}\label{eq:ripleys_O}
\end{equation}
Here, $\mathcal{S}_{O}(\mathbf{x}_{i},r,\triangle r)$ is defined
as the edge-corrected shell-like neighborhood around the particle
located at $\mathrm{\mathbf{x}}_{i}$, which is obtained by the intersection
of the VOI and the spherical shell or radius $r$ and thickness $\triangle r$
centered at the particle location (Fig. \ref{fig:2d-schemes}B):
\begin{equation}
\mathcal{S}_{O}(\mathbf{x_{\mathit{i}}},r,\triangle r)=\left\{ \forall\mathbf{x}\in\mathcal{V}|r-\triangle r/2\leq d(\mathbf{x},\mathbf{x}_{i})\leq r+\triangle r/2\right\} 
\end{equation}
Similarly, in Eq \ref{eq:ripleys_O}, $C(\mathbf{x},\mathcal{S}_{O}(\mathbf{x}_{i},r,\triangle r))$
is the number of particles located in the shell $\mathcal{S}_{O}(\mathbf{x}_{i},r,\triangle r)$
and $\mathcal{\mathit{V}}(\mathcal{S}_{O}(\mathbf{x}_{i},r,\triangle r))$
is the volume of this shell. Therefore, our definitions of $\mathcal{S}_{O}(\mathbf{x}_{i},r)$
and $\mathcal{S}_{O}(\mathbf{x}_{i},r,\triangle r)$ provide edge
corrections for functions $L$ and $O$.

Despite their conceptual similarity, these functions show important
differences. Function $O(r)$ can be considered more precise because
it depends only on particle pairs having distance close to $r$, while
$K(r)$ and $L(r)$ receive contributions from all scales between
$0$ and $r$. However, because $O(r)$ can be understood as a derivative
of $K(r)$ it is more affected by noise, especially when the density
of particles is low, which limits its usefulness in practice.

\subsubsection{Bivariate analysis}

The bivariate PPA functions allow investigations of the relationships
between two different particle patterns, the reference pattern, $\mathbf{X^{\mathit{r}}}=\left\{ \mathbf{x}_{i}^{r}\in\mathcal{V}|i=(1,\ldots,n)\right\} $,
and the evaluation pattern, $\mathbf{X^{\mathit{e}}}=\left\{ \mathbf{x}_{i}^{e}\in\mathcal{V}|i=(1,\ldots,n_{e})\right\} $.

The distribution function $G$ for the bivariate analysis is based
on the nearest distances among particles between the two different
patterns, $p_{G}^{re}(r)$:

\begin{equation}
G^{re}(r)=\int_{0}^{r}p_{G}^{re}(r)dr
\end{equation}

\noindent where $p_{G}^{re}(r)$ are distances from all reference
particles to their nearest evaluation particles. There is no bivariate
counterpart for function $F$ and consequently neither for $J$.

The bivariate versions of the Ripley's second-order functions analysis
show the co-localization of the evaluation particles in respect to
the reference particles. These functions are very similar to their
monovariate counterparts (Eqs. \ref{eq:ripleys_L} and \ref{eq:ripleys_O}):
\begin{equation}
\begin{array}{cc}
K^{re}(r)=\frac{4\pi r^{3}}{3}\cdot\frac{\sum_{i=0}^{n}C^{e}(\mathbf{x}^{e},\mathcal{S}_{L}(\mathbf{x}_{i}^{r},r))}{\lambda^{e}\cdot\sum_{i=0}^{n}\mathcal{\mathit{V}}(\mathcal{S}_{L}(\mathbf{x}_{i}^{r},r))} & \;L^{re}(r)=\sqrt[3]{\frac{K^{re}(r)}{4\pi}}-r\end{array}
\end{equation}

\begin{equation}
O^{re}(r)=\frac{\sum_{i=0}^{n}C^{e}(\mathbf{x}^{e},\mathcal{S}_{O}(\mathbf{x}_{i}^{r},r,\triangle r))}{\sum_{i=0}^{n}\mathit{\mathcal{\mathit{V}}}(\mathcal{S}_{O}(\mathbf{x}_{i}^{r},r,\triangle r))}
\end{equation}

\noindent except that $C^{e}$ is the number of evaluation particles
$\mathbf{X^{\mathit{e}}}$ within the edge corrected neighborhood
or the shell of reference particles $\mathbf{X^{\mathit{r}}}$.

\subsubsection{\label{subsec:null_models}Null-models and statistical inference}

In order to determine the statistical significance of particle distribution
analysis obtained by the first and second order PPA functions, statistical
hypothesis testing methods are used to compare the experimental with
the null-model results. Consequently, when generating proper null-models
for cryo-ET data, the shape of the particles and cellular compartments
has to be considered.

The CSR model was previously defined for point particles located in
arbitrary regions \cite{Wiegand2004}. Here we use the Complete Spatial
Randomness with Volume exclusion model (CSRV) in 3D, which is an extension
of CSR that takes into consideration the 3D shape of particles by
imposing volume exclusion to avoid particle overlap. Statistical comparison
of experimental results with CSRV null-model allows discarding the
random particle distribution hypothesis (the null hypothesis), in
which case it can be concluded that the particle organization is controlled
by a structural process. Furthermore, this comparison can show whether
the experimental distribution is more clustered or more uniformly
distributed than CSRV and determine length scale(s) at which the differences
are found. A more detailed analysis may require a null-model specifically
designed for the actual experimental question \cite{Rosenzweig2017}.

To assess the statistical significance of the results obtained by
the first order PPA functions $G$ and $F$, the Kolmogorov-Smirnov
(K-S) test is used. It requires that the variable studied is continuous,
and it is applied to the PPA functions represented as the cumulative
frequency distribution functions \cite{Andrey2010} to determine $D_{n,m}$
in the following way:
\begin{equation}
D_{n,m}=\underset{r_{D}}{\sup}|\tilde{G_{n}}(r)-G_{m}(r)|
\end{equation}

\noindent $\tilde{G_{n}}(r)$ is $G$ or $F$ function, obtained from
experimental data comprising $n$ particles and $G_{m}(r)$ the CSRV
null-model function, obtained by simulating $m$ synthetic instances
of the null-model with the same number of particles and the same VOI
as the experimental dataset. The null hypothesis stating that experimental
and the simulated distributions are identical can be rejected with
the probability $1-\alpha$ if \cite{Hodges:1958}:
\begin{equation}
D_{n,m}>\sqrt{-\left(\frac{m+1}{2nm}\right)\ln\left(\frac{\alpha}{2}\right)}
\end{equation}
The K-S test determines a single significance value of a given first
order PPA function, by taking into account the entire range of distances
that forms its domain. When testing function $G$, the positive sign
of $\tilde{G_{n}}(r_{D})-G_{m}(r_{D})$ specifies that the experimental
particles are clustered and the negative sign signifies that they
are uniformly distributed. The interpretation is the opposite when
testing function $F$.

For the second order PPA functions, the null hypothesis can be tested
by ranking PPA results obtained for multiple null-model simulations
and constructing an interval of confidence (IC) as a function of distance
$r$:
\begin{equation}
IC(r)=\left[IC(r)^{-},\;IC(r)^{+}\right]=\left[L_{\alpha}(r),\;L_{1-\alpha}(r)\right]
\end{equation}

\noindent where $L_{\alpha}(r)$ is the value of the second order
PPA function under consideration that has the rank of $\alpha\cdot100\%$.
If an experimental PPA function falls outside of the interval $IC(r)$
at the distance $r$, we can reject the null hypothesis fwith a confidence
$1-\alpha$. Additionally, if the experimental estimator is greater
(smaller) than $IC(r)^{+}$, we can conclude that the pattern is more
clustered (more uniformly distributed) than the null-model.

\subsection{\label{sec:impl}Methods}

\subsubsection{Computation of the second-order PPA functions}

Efficient implementations of many linear algebra operations that are
required to compute the PPA functions introduced above are already
available in libraries written in different programming languages.
Here we present implementations of the operations that are currently
not available in the standard libraries.

The central part of the second-order function computations concerns
the determination of the number of particles $C(\cdot)$ and the volume
$V(\cdot)$ of local edge corrected neighborhoods $\mathcal{S}_{L}$
and $\mathcal{S}_{O}$. Because the neighborhoods are defined in a
VOI of arbitrary geometry $\mathcal{V}$, these computations cannot
rely on closed formulas but require numerical approximations. Therefore,
they require computations of kernels for the number of particles and
volume for neighborhoods of different radii, for each particle $\mathbf{x}_{i}$
separately: $C(\mathbf{x},\mathcal{S}_{L}(\mathbf{x}_{i},r))$, $C(\mathbf{x},\mathcal{S}_{O}(\mathbf{x}_{i},r,\triangle r))$,
$V(\mathcal{S}_{L}(\mathbf{x}_{i},r))$ and $V(\mathcal{S}_{O}(\mathbf{x}_{i},r,\triangle r))$.
As a consequence, second-order analysis is computationally much more
demanding than first-order analysis. We note that the global particle
density can be computed by $\lambda=n/V(\mathcal{V})$.

\subsubsection{\label{subsec:Counting-the-number}Counting the number of particles
in a region of arbitrary shape}

It is critical to choose the most efficient way of evaluating whether
a point belongs to a VOI (${\ensuremath{\mathbf{x}\in\mathcal{V}}}$)
because this condition is evaluated intensively during the computation
of second-order functions. It is required both for the computations
involving real data and the randomly generated data used for null-models.
Here we consider two approaches that use different ways to represent
VOI.

\begin{figure}
\begin{centering}
\includegraphics[width=1\textwidth]{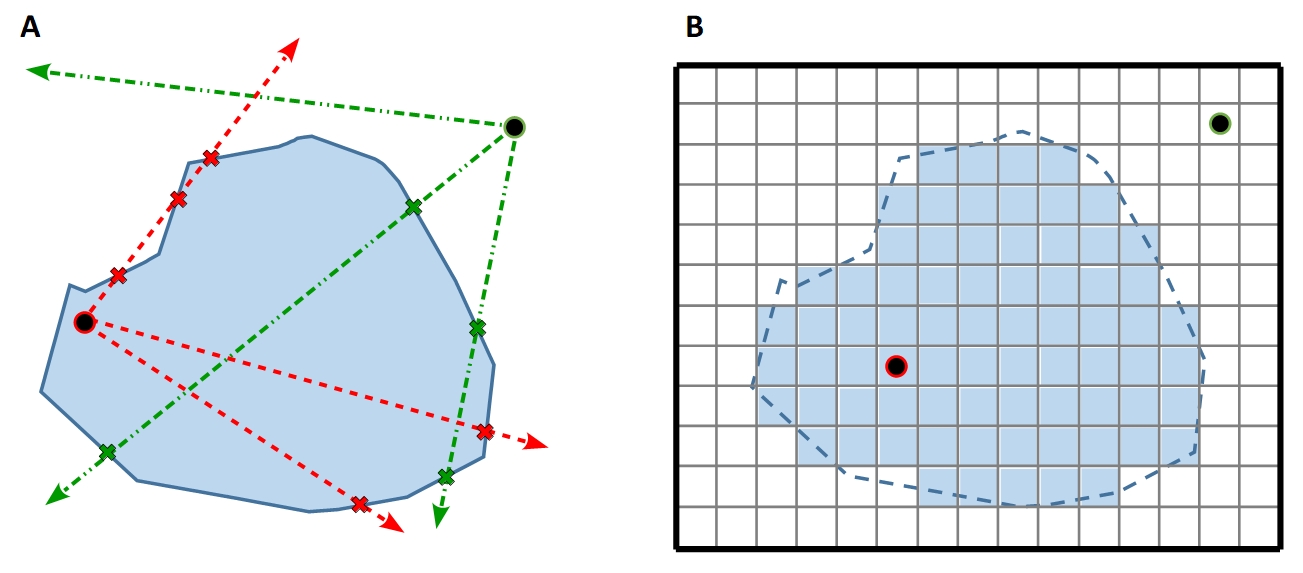}
\par\end{centering}
\caption{\label{fig:shc-num}Counting particles in a region of arbitrary shape.
A) The surface approach: VOI (blue field) is stored as a triangle
mesh (blue line), and the number of intersections (\textquotedblleft x\textquotedblright{}
symbols\textquotedblright ) between rays (dashed lines) and the VOI
surface are counted. B) The 3D array approach: VOI is represented
as \textit{true} voxels (blue field). VOI boundary is shown as dashed
blue line.}
\end{figure}

In the surface approach, VOI is represented as a closed surface defined
by a triangle mesh. To evaluate the condition ${\ensuremath{\mathbf{x}_{i}\in\mathcal{V}}}$,
where $\mathbf{x}_{i}$ is the particle center, we used the ray-firing
method consisting of the following steps: (1) Randomly oriented rays
originating at the evaluation point are generated. (2) For each ray,
the number of intersections with the bounding surface is counted.
(3) If the number of rays having an odd number of intersections is
larger than the number of rays having an even umber of intersections,
the point is considered to belong to the VOI (Fig. \ref{fig:shc-num}A).

In the second approach, a VOI is represented by \textit{true} voxels
in a dense 3D Boolean array (termed 3D array representation). A point
belongs to a VOI simply if it belongs to the set of \textit{true}
voxels (Fig. \ref{fig:shc-num}B).

It is clear that the surface approach uses less memory than the 3D
array approach, even though the array is binary. However, checking
the condition ${\ensuremath{\mathbf{x}\in\mathcal{V}}}$ for the 3D
arrays approach has the complexity $O(1)$, while in the surface approach
the complexity is $O(N_{rays})$, where $N_{rays}$ is the number
of rays used by the iterative ray-firing method \cite{Schroeder2006}.
Our results showed that for both approaches the running time per particle
decreased with the number of particles (Fig. \ref{fig:tm-num}). This
is likely a consequence of the hierarchical cache architecture of
current processors. Importantly, for up to 1000 particles, the running
time was smaller for the 3D array approach. Only for more than 1000
particles did the running time of the surface approach became comparable
to that of the 3D array approach. In addition, the running times were
similar for the cases when particles were randomly distributed in
a sphere or on a spherical shell.

\subsubsection{Estimating the volume of particle neighborhoods}

To calculate the volume of an edge corrected neighborhood $V(\mathcal{S})$,
we propose two different algorithms. These parallel the approaches
to determine whether a point belongs to a region introduced in the
previous section.

\begin{figure}
\begin{centering}
\includegraphics[width=1\textwidth]{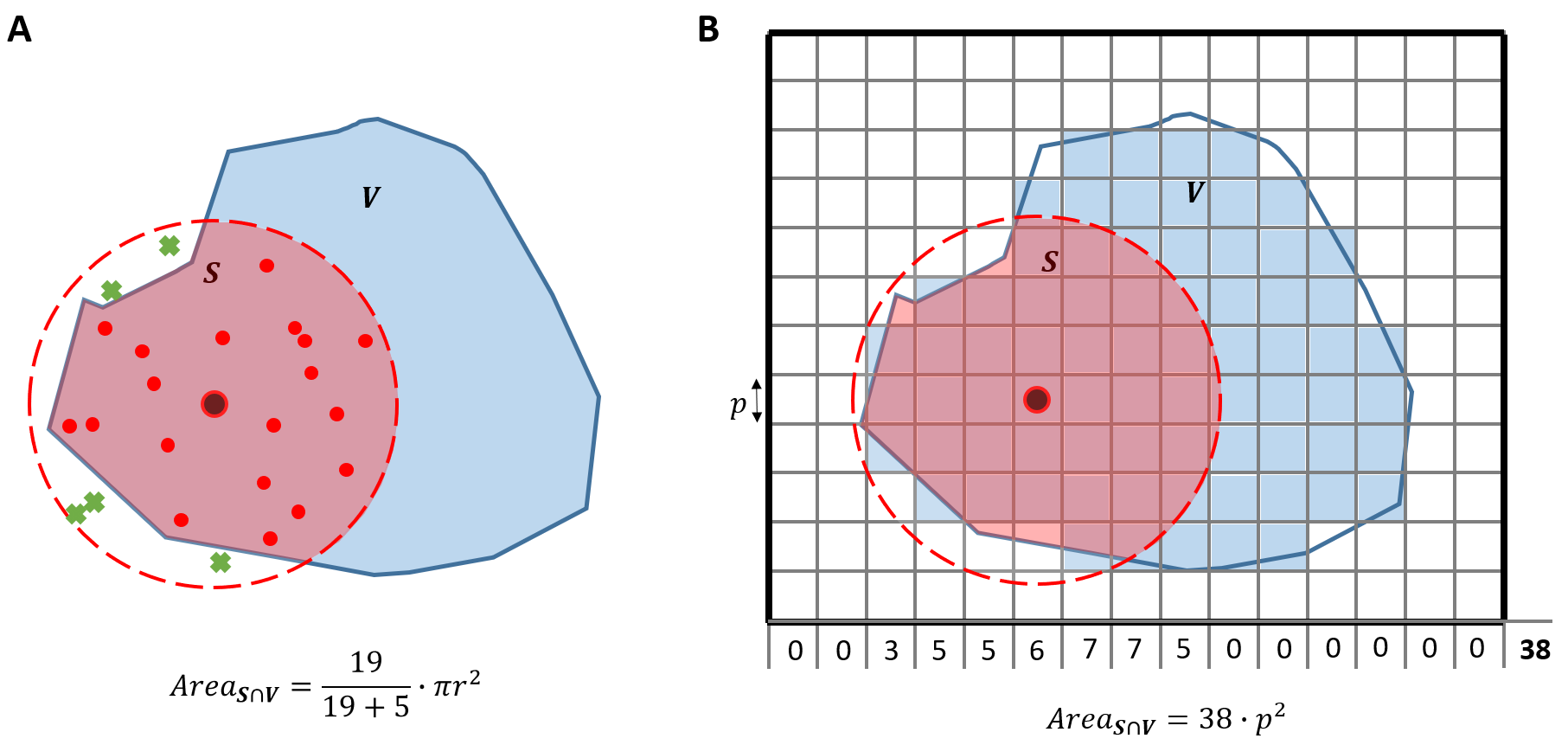}
\par\end{centering}
\caption{\label{fig:vol}2D schematics for irregularly bounded area computation.
A) MCS algorithm. B) 3D array method.}
\end{figure}

Monte Carlo surface algorithm (MCS) requires the surface VOI representation.
The volume of an edge-corrected neighborhood is calculated by randomly
generating points within a spherical neighborhood, using the surface
approach to determine whether these points belongs to VOI and counting
the number of points inside the VOI (Alg. \ref{alg:mc_volume}, Fig.
\ref{fig:vol}A).

The second, the 3D array algorithm, requires the 3D array VOI representation.
The volume of an edge-corrected neighborhood is determined by counting
the number of voxels that belong to both the spherical neighborhood
and the VOI using the array approach (Alg. \ref{alg:dsa}, Fig. \ref{fig:vol}B).
The 3D array used here has to be large enough to hold the entire neighborhood
$\mathcal{S}$ for the largest neighbor radius $r$. To speed up the
computations, the distance between each particle and all other voxels
is pre-computed, so that a single distance calculation for a particle
can be used for all neighborhoods of that particle.

To evaluate the precision of the neighborhood volume estimation by
MCS and 3D array algorithms, we applied them to particles located
at the distance of 5 nm to the boundary of a rectangular VOI, and
also calculated the edge corrected neighborhood volumes analytically.
The volume estimation error $E$ is defined as:

\begin{equation}
E[\%]=\frac{\hat{V}-V}{V}\cdot100
\end{equation}

\noindent where $V$ is the ground truth volume (computed analytically)
and $\hat{V}$ the estimated volume.

Our data showed that for both spherical and shell edge corrected neighborhoods,
10 - 1000 points and 10 rays per point the MCS algorithm performed
well for all neighborhood radii (Fig. \ref{fig:pr-MCS}). As expected,
increasing the number of points decreased the variability.

The 3D array algorithm was also very precise, with noticeable errors
only for very small neighborhoods, (Fig. \ref{fig:pr-DSA}). This
algorithm is expected to be less precise for small and non-flat neighborhoods
because in such conditions the discretization of space by voxels deviates
from the most the real (curved) shapes.

In comparison, the MCS algorithm required around 1000 particles to
reach the precision of the 3D array algorithm. Another advantage of
the 3D array algorithm is its non-stochastic nature.

MCS required a larger running time for neighbor size up to 75 voxels
and was comparable to 3D array for larger sizes (Fig \ref{fig:tm-MCS}).
Here we also used 10 rays per point for MCS. In both algorithms, the
running time per volume or area unit decreased with the increased
neighborhood size.

\subsubsection{Particle overlap}

Particles encountered in cryo-ET are not points but have a finite,
possibly complex 3D shape. Therefore when a new particle is added
to a synthetic null-model instance, it is necessary to ensure that
the particle does not overlap with any other particle. Here we use
the VTK library to implement volume exclusion between particles. Namely,
we first use the VTK library to generate the particle surface at the
intended position and orientation specified by Euler angles. Next,
we check whether the bounding box of the particle overlaps with any
of the other particle bounding boxes. If it does, we use again the
iterative ray-firing method implemented in VTK to ensure that no point
of a particle surface is inside the surface of any previously inserted
particle. If the particles indeed overlap, the new particle is rejected.

\subsubsection{Parallel implementation}

Second-order metric implementations are very CPU intensive. They require
long running times for real data comprising dozens of tomograms and
thousands of particles each. Moreover, to achieve a sufficient statistical
confidence, more than 100 simulations per tomogram should be computed.
To solve both problems, we provide a parallel implementation based
on the \textbf{\small{}\href{https://docs.python.org/2/library/multiprocessing.html}{multiprocessing}}
package of the Python programming language, which uses a shared memory
environment and exploits the internal parallelism of current multi-core
processors.

Specifically, particles are processed separately by $n$ independent
\textit{computational units. }To compute the second-order functions
for real data, these units are evenly distributed over $p\leq n$
processes that share access to VOI. For each particle, the distance
map (Euclidean or geodesic, see Section \ref{subsec:distance}), is
computed once and used for all distances, which minimizes the number
of times a distance map has to be computed (Fig. \ref{fig:sch_nhood}A,
B). For simulated data, the workload is divided by the number of tomograms
to be simulated ($m$) which are executed by $p\leq m$ processes
(Fig. \ref{fig:sch_nhood}C).

The computational speedup \cite{Amdahl1967} obtained by parallelization
was almost linear up to 15 processes and continued to increase until
the maximum number of processes (35) for both real and simulated data
(Fig. \ref{fig:sch_nhood}D). For a high number of processes, the
speedup was a little higher for the simulated data. This is beneficial
for our applications because most of the workload is generated by
the analysis of the null-model. Five synthetic tomograms with 500x500x100
voxels and 200 particles each were used for second-order function
computations. To ensure a fair comparison, the number of simulated
tomograms on each instance was the same as the number of concurrent
processes.

\subsubsection{Computational requirements}

This software package has been developed in Python and is available
open-source (see Code availability), in order to facilitate its dissemination
in the research community and the development of future extensions.
Graphs are plotted using \textit{matplotlib} library \cite{Hunter2007},
and surface meshes are stored and processed using VTK \cite{Schroeder2006}
and visualized through Paraview \cite{Ayachit2015}. All computing
experiments were executed on a computer node of the Max Planck Institute
of Biochemistry cluster, it has 500GB RAM with 36 processors Intel(R)
Xeon(R) CPU E5-2699 v3 @ 2.30GHz with SUSE Linux Enterprise Server
12 SLES 12 SP 1 Operation System.

\section{Results}

\subsection{Validation with synthetic data}

\subsubsection{Uni- and bivariate functions}

\begin{figure}
\begin{centering}
\includegraphics[width=1\textwidth]{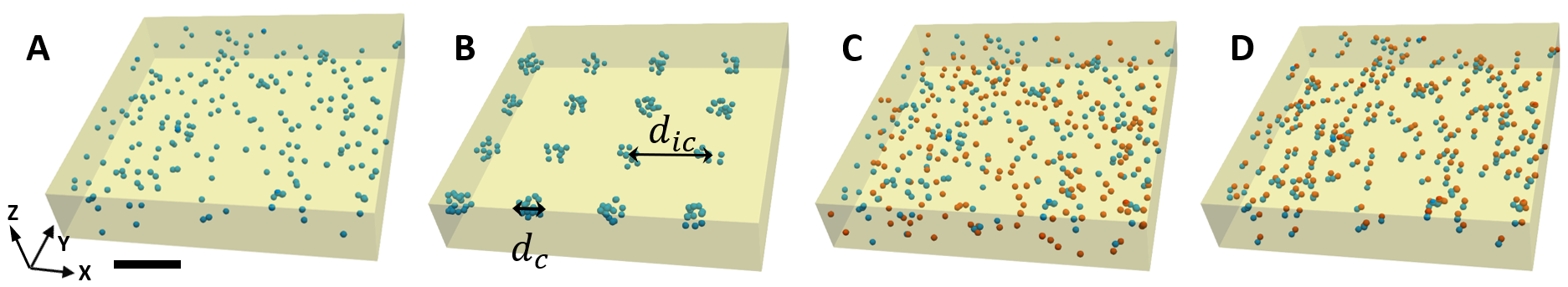}
\par\end{centering}
\caption{\label{fig:syn-tomos}Synthetic datasets used for validations. A)
CSRV. B) SRPV with cluster diameter $d_{c}\approx46$ voxels and intercluster
distance $d_{ic}=125$. C) Two independently generated CSRV patterns
(blue and red). D) CSRV pattern (blue), points of the spatially correlated
pattern (red) are located at distances obtained from the normal distribution
$\mathcal{N}(\mu=10,\sigma=1)$ to the CSRV pattern, we chose the
parameters of the normal distribution so that the two patterns visually
colocalize. In all cases particles (blue and red) were spheres of
radius $r_{p}=5$ each particle pattern contained 200 particles and
tomograms had size 500x500x100 voxels. Scale bar 100 voxels.}
\end{figure}

To validate the implementation of the second-order metrics, we generated
synthetic tomograms that contain randomly distributed and clustered
particles (Fig. \ref{fig:syn-tomos}A, B). Particle clusters were
generated using Sinusoidal Random Pattern (SRP) distribution $\tilde{U}$:
\begin{equation}
\tilde{U}\left((x=q\pi x',y=q\pi y',z=q\pi z')|\sin x+\sin y+\sin z>3t\right)
\end{equation}
where $x'$, $y'$ and $z'$ are random real numbers uniformly distributed
in the interval $[-1,1]$, $q\in N$ and $t\in[0,1]$. Distribution
$\tilde{U}$ creates clusters that have oval shape of maximal radius
$r_{c}\approx\left(q\pi\right)^{-1}arccos(3t-2)$ (in $x'$, $y'$,
$z'$ units). The clusters reach an almost spherical shape for $t>0.8$,
in which case the radius can be approximated by $r_{c}\approx\left(q\pi\right)^{-1}\sqrt{6(1-t)}$.
The distance between neighboring clusters (center to center) is $d_{ic}=\nicefrac{2}{q}$
(in $x'$, $y'$, $z'$ units).

Here we generated multiple CSRV, and SRP with volume exclusion (SRPV)
synthetic datasets, where SRPV was created from SRP populated with
finite size particles (as opposed to point-particles) and excluding
overlapping particles. Our software allows using any closed surface
to represent particles. For simplicity, here we used spheres with
radius $r_{p}=5$. This synthetic datasets had a size of 500x500x100
voxels, where 500 voxels corresponds to $x'$, $y'$, and $z'$interval
$[-1,1]$. For SRPV, we set $q=4$, and $t=0.8$, thus generating
16 clusters at the inter-cluster distance of $d_{ic}=125$ voxels
(Fig. \ref{fig:syn-tomos}B).

\begin{figure}
\begin{centering}
\includegraphics[width=1\textwidth]{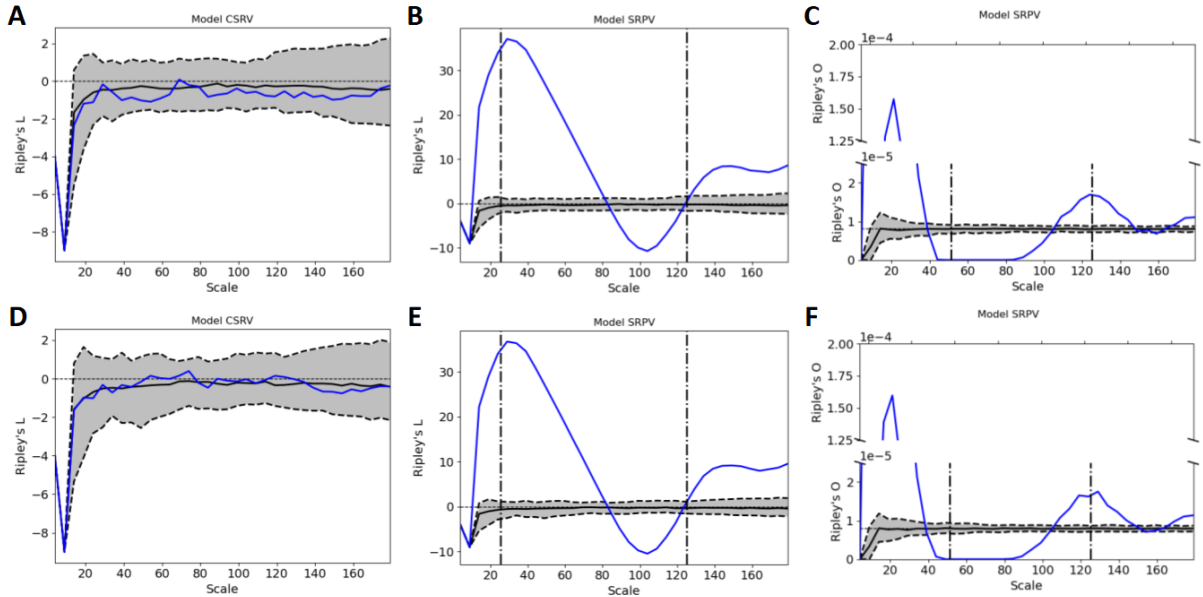}
\par\end{centering}
\caption{\label{fig:2nd-syntetic}Univariate second-order metrics validation
with synthetic datasets. (A-C) 3D array and (D-F) MCS algorithm. (A,
D) Function $L$ for CSRV model, blue line shows the mean of additional
five CSRV simulations. (B, E) Function $L$ for SRPV model, shown
are the mean of five simulations (blue line) and the CSRV null-model
(grey). (C, F) Function $O$ for SPRV model, shown are the mean of
five simulations (blue line) and the CSRV null-model (grey). In all
cases the grey area and line show $IC\:5-95\%$ and the mean of 100
CSRV simulations, respectively. For MCS the number of iterations for
convergence is $1000$ and the maximum $100000$. In (B, E) the vertical
dashed lines mark $d_{c}/2$ and $d_{ic}$ , and in (C, F) $d_{c}$
and $d_{ic}$ respectively. In (A, B, D, E) the horizontal line marks
$X=0$, and in (C, F) the global density $\lambda$, in all cases
these lines represent the behavior of the ideal random pattern.}
\end{figure}

To validate our implementation of the univariate Ripley's functions,
we first checked the $L$ function obtained for the CSRV dataset (the
null-model). The mean value of 100 simulations was close to $0$ and
the $IC\:5-95\%$ was spread around the $0$ value, as expected for
a pure random pattern (Fig. \ref{fig:2nd-syntetic}A, D). This was
the case for the whole distance range except for distances lower than
$2r_{p}$, indicating that at short distances particle volume exclusion
dominates. We also calculated the $L$ function for additional five
CSRV simulations, their mean shows the stochasticity expected for
the analysis of real randomly distributed particles.

The univariate Ripley's functions that we obtained for 100 SRPV simulations
did not differ between MCS and the 3D array algorithms, Next, we calculated
the univariate Ripley's functions $L$ and $O$ functions for the
clustered, SRPV, datasets (Fig. \ref{fig:2nd-syntetic}B, C, E, F).
Using the MCS and the 3D array algorithms produced virtually indistinguishable
results, as was the case for the CSRV null-model (compare panels A-C
and D-F of Fig. \ref{fig:2nd-syntetic}). The calculated $L$ and
$O$ functions for SRPV model showed significant deviations from the
results obtained for the CSRV null-model. Specifically, the distance
at which function $L$ reached maximum and function $O$ minimum were
close to $d_{c}/2$ and $d_{c}$ respectively, as expected based on
analytical calculations \cite{Kiskowski2009}. Furthermore, function
$L$ was zero and its first derivative was positive at the distance
that corresponds to the inter-cluster distance. This can be explained
by observing that the concentration of particles in a neighborhood
of radius that equals the periodicity of the point pattern is simply
the global particle concentration. At the same distance, function
$O$ reached a local maximum, which directly points to clustering
at that distance $d_{ic}$.

\begin{figure}
\begin{centering}
\includegraphics[width=1\textwidth]{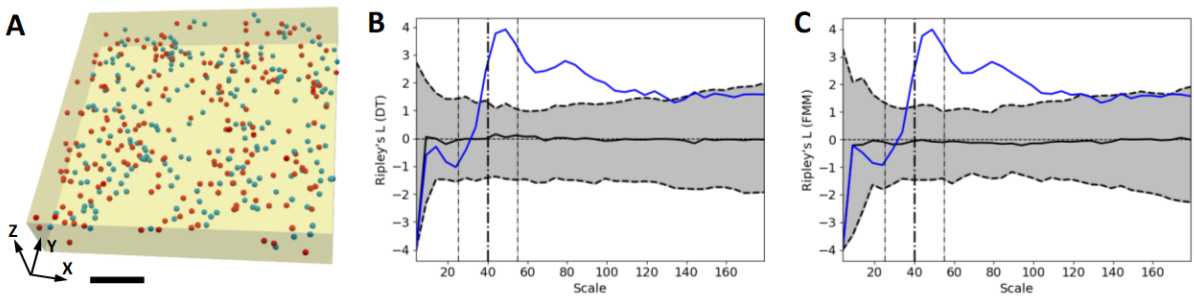}
\par\end{centering}
\caption{\label{fig:2nd_bi}Bivariate second-order analysis validation. (A)
Two correlated point patterns where each particle of pattern 2 (red)
are placed at a distance controlled by a distribution $\mathcal{N}(\mu=40,\sigma=5)$
to a particle of pattern 1 (blue). (B) Function L for the correlated
pattern shown in A (mean of 5 simulations) is shown in blue, and the
$IC\:5-95\%$ of 100 simulated uncorrelated pairs in gray. Distances
were measured using Distance Transform (DT). (C) Like B except that
Fast Marching Method (FMM) was used for distance computation. (B,
C) Thick vertical lines represent $r=\mu$ and the thin ones $r=\mu\pm3\sigma$.
Tomograms size 500x500x100. Scale bar 100 voxels.}
\end{figure}

Finally, to validate bivariate Ripley's functions, we generated datasets
containing two particle sets, the particle sets were spatially uncorrelated
in some and spatially correlated with each other in other datasets.
The uncorrelated datasets consisted of two independent CSRV patterns
(Fig \ref{fig:syn-tomos}C). To generate the correlated datasets,
we first generated a CSRV pattern and then placed particles of the
second set at distances following the Normal distribution $\mathcal{N}(\mu,\sigma)$
from randomly selected CSRV particles (Figs. \ref{fig:syn-tomos}D
and \ref{fig:2nd_bi}.A). For both MCS and the 3D array algorithms,
the mean value of the bivariate $L$ function for 100 simulations
of the null-model was very close to 0 at all distances, except for
the very short ones where the effects of volume exclusion dominate,
thus showing that the two sets were independent (Fig. \ref{fig:2nd_bi}B).
For the correlated datasets, the bivariate $L$ function of the correlated
dataset was significantly different from those obtained for the uncorrelated
null-model at an intermediate range of distances. The bivariate $L$
function for the correlated dataset reached significance at the distance
of approximately $\mu-\sigma$ and reached a maximum at $\mu+\sigma$,
which agrees with the criterion customarily used in the field \cite{Kiskowski2009}.

All together, our implementation of the second-order uni- and bivariate
functions on synthetic datasets yielded results that allowed the determination
of the correct clustering distance scales. Furthermore, the MCS and
the 3D array based implementations produced almost identical results,
thus validating both algorithms.

\subsubsection{\label{subsec:distance}Influence of the distance metric}

Because a VOI can have an arbitrary 3D shape and it may include holes,
the choice of distance metric $d$ used for defining the neighborhoods
$\mathcal{S}_{L}$ and $\mathcal{S}_{O}$, Euclidean or geodesic,
can have a strong impact on the computation of the second-order functions.
We used the Distance Transform (DT) \cite{Ragnemalm1993} for computing
Euclidean and Fast Marching Method (FMM) algorithm \cite{Sethian1996}
for calculating geodesic distance. We implemented FMM only for the
3D array VOI representation.

For a VOI that has convex shape and trivial topology in 3D (such as
that shown on Fig. \ref{fig:2nd_bi}A), Euclidean and geodesic distances
are the same, resulting in essentially the same $L$ and $O$ functions
(Fig. \ref{fig:2nd_bi}B, C).

\begin{figure}
\begin{centering}
\includegraphics[width=1\textwidth]{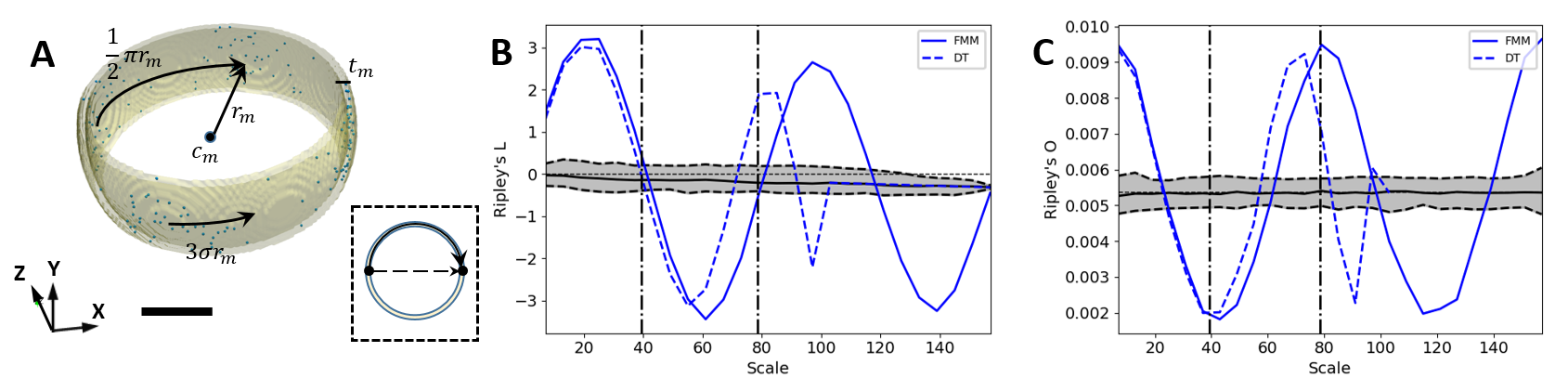}
\par\end{centering}
\caption{\label{fig:dst}Geodesic and Euclidean distance based second order
functions on a complex VOI. (A) A biological membrane-like VOI, $r_{m}=50$,
and $\sigma=0.25$ containing clustered particles (blue points). The
inset shows the difference between geodesic (solid arrow) and Euclidean
distances (dashed arrow) between two points on the opposite sides
of a sphere. (B) Function $L$ and (C) function $O$ for the clustered
particle pattern (blue lines) and the null-model ($IC\:5-95\%$ in
gray, black line represents the mean). In both cases the second order
functions based on geodesic distance (solid and blue line) and Euclidean
distance (dashed blue line) are shown. Scale bar 30 voxels.}
\end{figure}

However, cellular VOIs often have a complex shape because they are
delineated by biological membranes or are formed by a particular distribution
of molecular complexes. To evaluate the PPA functions obtained using
the two distance metrics, we generated a synthetic dataset where VOI
takes the shape of a cropped spherical shell of radius $r_{m}=50$
and thickness $t_{m}=6$ voxels (Fig. \ref{fig:dst}A). Such VOIs
are encountered for complexes bound to a membrane of a cellular organelle
\cite{Martinez2020}. The particle pattern was formed by four clusters
localized in the VOI, and for each cluster $i$, particles were distributed
according to the following expression:
\begin{equation}
\tilde{U_{i}}\left(x=x_{c}+r_{m}\cos\left(\phi+\varphi_{i}\right)\sin\phi,y=y_{c}+r_{m}\sin\left(\phi+\varphi_{i}\right)\sin\phi,z=z_{c}+r_{m}\cos\theta\right)
\end{equation}
where $c_{m}=(x_{c},y_{c},z_{c})$ are the coordinates of the VOI,
$\mathcal{N}(\mu=0,\sigma=0.25)$, $r_{m}$ is the radius of the VOI
$\theta$ and $\phi$ are spherical angles taken from the normal distribution
and $\varphi_{1,2,3,4}=\left\{ 0,\nicefrac{\pi}{2},\pi,\nicefrac{3\pi}{2}\right\} $
define cluster centers. In this way, the center-to-center inter-cluster
distance was $\frac{1}{2}\pi r_{m}$ (78.5 voxels for $r_{m}$ of
50~nm) and the cluster size can be approximated by $3\sigma r_{m}$
(37.5 voxels). In addition, particle distribution $\tilde{U}(x,y,z)$
takes volume exclusion into account.

Our results showed that functions $L$ and $O$ were significantly
different from the CSRV null-model. The geodesic distance based $L$
function had the first maximum at approximately 20 voxels ($1.6\sigma r_{m}$),
a 0 crossing with a negative slope at 40 voxels and another with a
positive slope at 80 voxels, while the geodesic $O$ function had
a local maximum at 80 voxels (Fig. \ref{fig:dst}B, C). These correspond
to the size of the clusters and the distance between them, in the
same way that functions $L$ and $O$ did for a simple VOI (Figs \ref{fig:syn-tomos}A
and \ref{fig:2nd-syntetic}B, C, E, F). However, the Euclidean distance
based $L$ and $O$ functions showed a shift towards shorter distances.
Additionally, the geodesic distance based $O$ function of the null-model
particle pattern (CSRV), was almost perfectly flat, as expected for
the ideal case (Fig. \ref{fig:dst}C). The $L$ function for the null-model
deviated slightly from the expected 0-value at the largest distances,
corresponding to the neighborhoods that reach to the opposite side
of the VOI (Fig. \ref{fig:dst}B). This is likely because the geodesic
distance between some of the distant points on the cropped spherical
shell VOI differs from the geodesic distance on a full spherical shell.
Therefore, our results show that using FMM allows an accurate determination
of the geodesic distance and that for membrane-like compartments and
other complex shape VOIs, the PPA functions should be calculated based
on geodesic and not Euclidean distances for .

\subsection{\label{subsec:exp}Experimental data}

\begin{figure}
\begin{centering}
\includegraphics[width=0.95\textwidth]{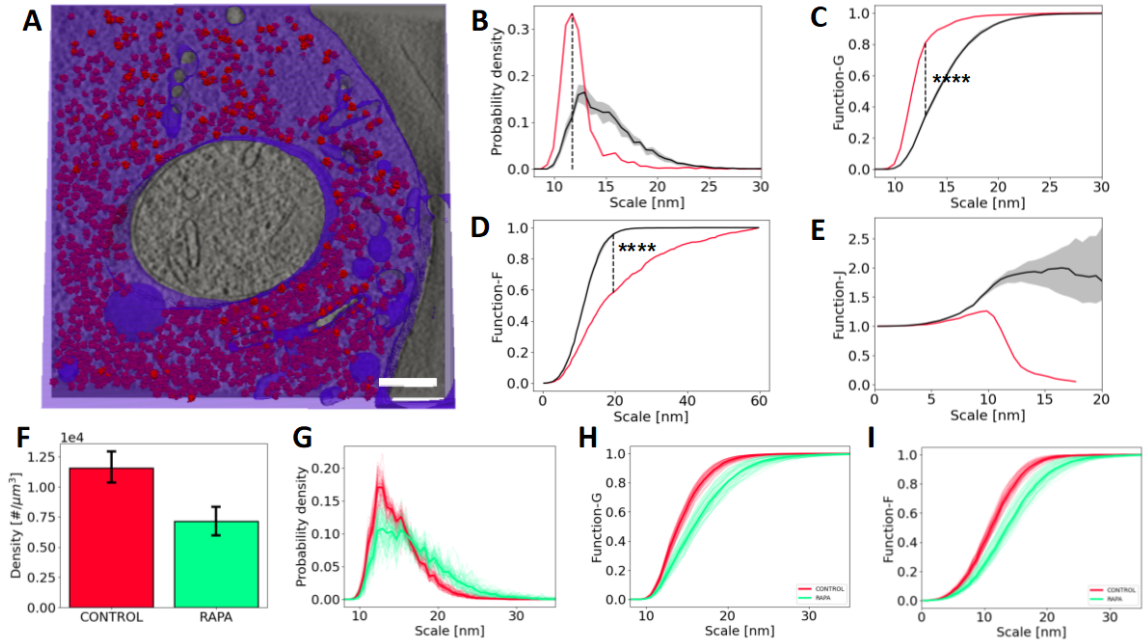}
\par\end{centering}
\caption{\label{fig:exp_1st}First-order analysis of experimental data. A)
Tomographic slice of the analyzed cryo-tomograms, with overlay showing
segmented cytoplasm (VOI, transparent blue) and localized ribosomes
(red), scale bar 200 nm. B) Histogram of all nearest neighbor distances
(function $G$), dashed vertical line marks the most frequent nearest
neighbor distance. C) Function $G$ shown as a cumulative distribution
(K-S test, $D_{2182,20}=0.4703$ and $\alpha<0.0001$). D) Spatial
contact distribution function ($F$) with 1000 simulated points (K-S
test, $D_{1000,20}=-0.3743$ and $\alpha<0.0001$). E) Function $J$.
B-E) The functions obtained from the experimental data are shown in
red. CSRV null-model simulation means are shown in black and $IC\:5-95\%$
in grey (20 simulations for each tomogram). Dashed vertical lines
show $D_{n,m}$. F-I) Comparison between the two sets of tomograms,
control (CONTROL, red), and Rapamycin treated (RAPA, green). F) Particle
density within the VOIs, mean and $IC\:5-95\%$. G) Function $G$
shown as histogram. H) Function $G$ shown as the cumulative distribution.
I) Function $F$. G-I) Analyses of individual simulated null-models
are shown as thin lines, thick lines show the means. In all cases
Euclidean distance was used.}
\end{figure}

We proceeded to validate the computational methods described in previous
sections on experimental data. To this end, we applied the first and
second order PPA functions to study the organization of ribosomes
in yeast cells visualized by cryo-ET. We used data form two experimental
groups, each comprising a set of tomograms from yeast cell cytoplasm
imaged \textit{in situ}. The first set contains 13 tomograms of yeast
cells that were treated with Rapamycin and the second 14 tomograms
of untreated cells \cite{Delarue2018}. In both cases, cell cytoplasm
was segmented, ribosomes were localized by template matching and a
high resolution structure of ribosomes (obtained from EMD-3068) was
used to represent the ribosome shape, as shown on a slice of a labeled
tomogram (Fig. \ref{fig:exp_1st}A).

We calculated the first-order functions for each tomogram separately
in order to determine the lengths that characterize the organization
of ribosomes (Fig. \ref{fig:exp_1st}.B-E). In untreated cells, the
nearest neighbor function showed a clear peak that represents the
most commonnearest-ribosome distance. For each tomogram, multiple
null-model distributions were generated, each having the same VOI
and the same number of ribosomes as the corresponding tomogram. All
functions ($G$, $F$, and $J$) showed that ribosomes are significantly
clustered with respect to the null-model (CSRV). This is in agreement
with the expected aggregation of functional ribosomes to form poly-ribosomes.
More generally, this type of analysis can help describing macromolecular
interactions \cite{Rosenzweig2017,Jasnin2019}. However, it is not
straightforward to combine the first order functions of individual
tomograms within an experimental group to obtain a single function
that characterizes an experimental condition, because the particle
concentration within VOIs differ between the cells even though they
were grown under identical conditions, which influences the first
order functions (Fig. \ref{fig:exp_1st}F-I).

\begin{figure}
\begin{centering}
\includegraphics[width=0.95\textwidth]{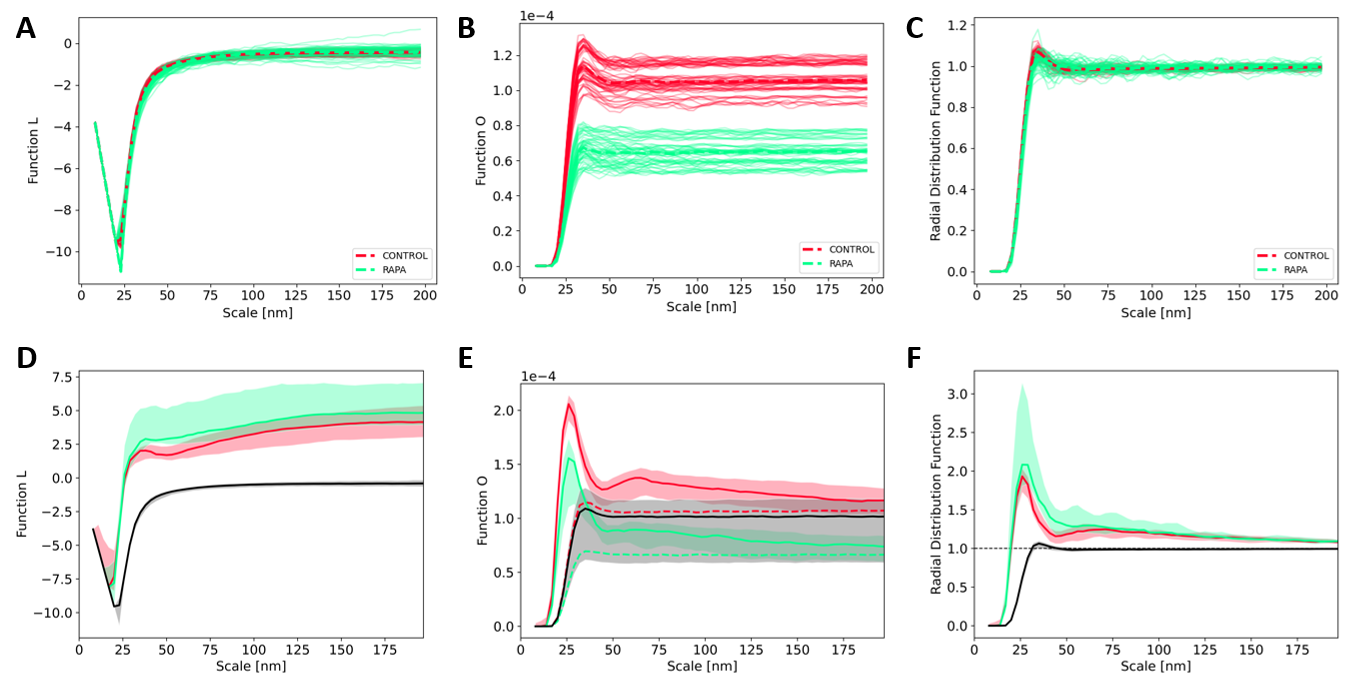}
\par\end{centering}
\caption{\label{fig:exp_2nd}Second-order analysis for experimental and the
corresponding CSRV null-model data. A-C) Functions obtained for CSRV
null-model data corresponding to the control (CONTROL, red) and Rapamycin
treated (RAPA, green) tomograms. Each thin semi-transparent line represents
a CSRV null-model simulation, five per tomogram, the thick dashed
lines show the means. D-F) Functions obtained for experimental tomograms,
control (CONTROL, red) and Rapamycin treated (RAPA, green). Pooled
null-model simulations (untreated and Rapamycin treated) $IC\:5-95\%$
is shown in grey and the mean in black. E) Also shows the mean simulations
of the experimental groups (dashed lines). A, D) Function $L$. B,
E) Function $O$. C, F) Radial distribution function.}
\end{figure}

Second-order PPA functions were calculated for each tomogram separately,
as well as for their respective CSRV null-models (Fig. \ref{fig:exp_2nd}).
The null-model $L$ functions for different tomograms and different
experimental groups (control and Rapamycin treated) were very similar.
The experimental group means were almost indistinguishable, only the
variability was higher for the Rapamycin set likely because of the
lower number of particles (Fig. \ref{fig:exp_2nd}.A). In contrast,
function $O$ for the null-model was very different for different
tomograms and there was a clear separation between the two experimental
groups. Because this variability was likely due to the different global
particle concentration in the tomograms, we here propose to use the
radial distribution function $g(r)$ \cite{Chandler1987}, which is
computed by normalizing function $O$ by the global particle concentration:

\begin{equation}
g(r)=\frac{1}{\lambda}O(r)
\end{equation}

\noindent This normalization restored the low variability of the null-model
simulations between tomograms and experimental groups (see Fig. \ref{fig:exp_2nd}.C).

When applied to the experimental data, the $L$ and the radial distribution
functions clearly showed significant clustering, both for all tomograms
taken together and for the experimental groups taken separately (Fig.
\ref{fig:exp_2nd}.D, F). In both experimental cases the first maximum
was located around 25 nm, which approximately corresponds to the double
of the most frequent nearest neighbor distance \ref{fig:exp_1st}.B).
While the large variability between the groups obtained for the function
$O$ precludes the interpretation of all tomograms taken together,
each of the experimental group showed significant clustering when
compared to the corresponding null-model (Fig. \ref{fig:exp_2nd}B,
E). While it is expected that in the untreated cells the formation
of polyribosomes leads to ribosome clustering, our results show that
some form of ribosome clustering persists in the Rapamycin treated
cells.

Furthermore, in order to set the stage for a statistical comparison
between the experimental groups, we computed the mean and $IC\:5-95\%$
as we did before, except that because of the low number of tomograms,
the $IC$ contains all tomograms. This comparison is valid for L and
radial distribution function as null-models converge to a similar
$IC$. However, in our case, this approach can not be applied to function
$O$, because we already saw that it is sensitive to particle global
densities and the two experimental groups contain different number
of particles. The statistical significance between the groups can
be easily established for distances at which the $L$ and the radial
distribution functions show clear separation between the experimental
groups. To determine the significance at other distances, non-parameteric
inference tests could be applied to values of the second order functions.

\section{Discussion}

Here we implemented the first and second-order, mono- and bivariate
PPA functions for 3D VOIs of arbitrary shape. This is of particular
importance for the analysis of molecular complexes visualized in 3D
biological images, such as those obtained by cryo-ET of cellular samples,
because cellular regions often have complex, non-convex shape. These
cellular regions can be be formed by lipid membrane-bounded organelles
and vesicules, membranes where certain membrane-bound complexes reside,
or any other cellular regions that constrain the localization of molecular
complexes of interest. Furthermore, because molecular complexes occupy
a substantial part of the cellular volume and the influence of their
size cannot be neglected at short distances, we represented molecular
complexes as 3D objects and imposed volume exclusion to prevent the
overlap.

The implementation of the second-order functions was particularly
important because they are sensitive to clustering at multiple distance
scales, thus providing more information than the first order functions.
It critically depends on solving two tasks; (1) determination of the
number of particles located within an irregularly bounded spatial
region and (2) measuring the volume of this space. The solutions depend
on the approach taken to represent spatial regions (particle neighborhoods
and VOI). In the first approach, we represented spatial regions by
a triangular surface mesh that defines the region boundary. Here we
used the stochastic ray-firing algorithm to count the number of particles
and MCS for the volume determination. The second approach we implemented
is based on representing spatial regions as 3D binary arrays, which
makes the implementation of the two tasks trivial. In theory, ray-firing
and MCS are stochastic methods that can achieve any precision and
are independent of the volume shape. 3D array based methods are deterministic
and they typically use more memory than the surface based methods.
However, our results show that for the same computational time, the
3D array-based methods achieved a higher precision than the surface-based
counterparts. Nevertheless, the second order PPA functions that we
computed using these two approaches were virtually indistinguishable.

We validated the implementation of the second-order, uni- and bi-variate
PPA functions on synthetic datasets. Our software correctly detected
the clustering distance scales that characterized the particle distribution
in the synthetic datasets. Additionally, we implemented the PPA functions
based on both Euclidean and geodesic distances, verified that they
produced the correct results, and showed that for simple VOIs the
Euclidean and the geodesic results were the same. Importantly, the
geodesic distance based second order PPA functions were more suitable
for applications involving complex VOIs, such as for describing the
nanodomain organization of molecular complexes located on curved membranes,
the situation commonly encountered in cellular environments.

Because analytical solutions for the PPA functions do not exist for
complex VOI and particle geometries, synthetic random tomograms (null-models)
are required in order to determine the statistical significance of
the PPA functions applied to experimental data. The inherent variability
of cellular components and the fact that a typical cellular cryo-ET
dataset contains tens of tomograms necessitate generating a set of
synthetic random model tomograms for each experimental tomograms.
An experimental tomogram and its corresponding synthetic tomograms
have to have the same VOI and the number of particles. Furthermore,
a large number of synthetic tomograms is needed to allow reaching
a specified significance level. To alleviate the computational burden
involved in the generation of the necessary number of random models
and the computationally intensive calculation of the PPA functions
for all tomograms, we propose a multi-process implementation of these
routines that reduced the running times, thus enabling the effective
analysis of realistic datasets. We achieved a speed-up factor of approximately
15 using a single processor multi-core architecture, a similar value
to those recently obtained in \cite{Wang2019} where a cluster of
computers was used for computing space-time Ripley's K.

Application of our software to yeast cell cytoplasm imaged by cryo-ET
allowed us to detect ribosome clustering and determine the characteristic
distance scales. Furthermore, we showed that the second order functions,
in particular $L$ and the radial distribution functions, were better
suited to compare experimental groups comprising multiple tomograms
because they were not sensitive to the global concentration of particles.

Future applications of our software are also expected to provide a
spatial characterization of macromolecular crowding, as well as liquid
and lipid phase separation. These processes recently gained a significant
biological interest, because they were shown to affect biochemical
reactions in cells and global organization of cellular membranes and
regions in different cellular systems \cite{Loewe_Kedrov:2020_rev,Chen_Zhang:2020_rev}.

Therefore, our implementation of the PPA functions provides a tool
that can characterize simultaneous clustering at multiple distance
scales, which is suitable for applications to cellular molecular complexes
visualized by cryo-ET, as well as to other 3D systems where real-size
particles are located within regions possessing complex geometry.

\section*{Acknowledgments}

We thank S. Pfeffer and B. Engel for giving us access to their cryo-ET
experimental data. This work was supported by the European Commission
(grant no. FP7 GA ERC-2012-SyG\_318987-- ToPAG) the Deutsche Forschungsgemeinschaft
(DFG, German Research Foundation) under Germany\textquoteright s Excellence
Strategy - EXC 2067/1- 390729940 and by Max Planck Society. We would
like to thank Gabriela J. Greif for critical reading of the manuscript.

\section*{Code availability}

PyOrg code belongs to a more general public repository (pyseg\_system,
\href{https://github.com/anmartinezs/pyseg_system}{https://github.com/anmartinezs/pyseg\_system})
which contains additional software, tests, and dependencies. The code
of PyOrg alone and can be downloaded from \href{https://github.com/anmartinezs/pyseg_system/tree/master/pyseg-sys/pyorg}{https://github.com/anmartinezs/pyseg\_system/tree/master/pyseg-sys/pyorg}.

\noindent \bibliographystyle{unsrt}
\bibliography{references}

\part*{\protect\pagebreak Supplementary material}

\renewcommand{\thefigure}{S\arabic{figure}}
\renewcommand{\thealgorithm}{S\arabic{algorithm}}
\setcounter{figure}{0}

\begin{algorithm}
\begin{algorithmic} 
 \Require{$c$: Particle center, $r$: Scale, $\mathcal{V}$: cellular compartment, $N$: Num. iterations to converge, $M$: Max. number of iterations}  
 \Ensure{$\hat{V}(\mathcal{S}(c,r))$: Volume for a particle neighborhood embedded in $\mathcal{V}$}
 \State{$N_{hits} \gets 0$, $N_{iter} \gets 0$}
 \While{$\left(N_{hits} \leq N\right) \wedge \left(N_{iter} \leq M\right)$} 
  \State{$N_{iter} \gets N_{iter}+1$}
  \State{$\textbf{x} \gets c + \mathrm{gen\_rand\_coordinate(r)}$}
   \Comment{For function L use the 3-ball algorithm, for function O the 2-sphere}
  \If{$\mathbf{x} \in \mathcal{V}$}
   \State{$N_{hits} \gets N_{hits}+1$}
  \EndIf
 \EndWhile
 \State{$\hat{V}(\mathcal{S}) = \left( N_{hits}/N_{iter} \right) \cdot V(\mathcal{S}(c))$}
\end{algorithmic}\caption{\label{alg:mc_volume}Computation of the volume of one particle edge
corrected neighborhood by MCS method.}
\end{algorithm}

\begin{algorithm}
\begin{algorithmic} 
 \Require{$r$: Neigbourhood scale}  
 \Ensure{$\mathbf{s}$: Random sample in the 2-sphere}
 \State{$\mathbf{x} \gets (\mathcal{N}(0, 1), \mathcal{N}(0, 1), \mathcal{N}(0, 1))$}
 \Comment{$\mathcal{N}(\mu, \sigma)$: is the normal random distribution with parameters $\mu$ and $\sigma$}
 \State{$\mathbf{s} = r \cdot \frac{\mathbf{x}}{\left \| \mathbf{x} \right \|}$}
\end{algorithmic}

\caption{Uniformly sampling the 2-sphere, adapted from the Müller-Box algorithm
\cite{Box1958}.}
\end{algorithm}

\begin{algorithm}
\begin{algorithmic} 
 \Require{$r$: Neigbourhood scale}  
 \Ensure{$\mathbf{s}$: Random sample in the 3-ball}
 \State{$\mathbf{x} \gets (\mathcal{N}(0, 1), \mathcal{N}(0, 1), \mathcal{N}(0, 1))$}
 \Comment{$\mathcal{U}$: is the uniform random distribution in $[0,1]$}
 \State{$\mathbf{s} = r \cdot \mathcal{U}^{1/3} \cdot \frac{\mathbf{x}}{\left \| \mathbf{x} \right \|}$}
\end{algorithmic}

\caption{Uniformly sampling the 3-ball, adapted from the Müller-Box algorithm
\cite{Box1958}.}
\end{algorithm}

\begin{algorithm}
\begin{algorithmic} 
 \Require{$\mathbf{c}$: Particle center, $\{r_1,...,r_n\}$: set of n scales, $\mathcal{V}$: (optional) Shell thickness $t$}  
 \Ensure{${\hat{V}_{1}(\mathcal{S}(c,r_1)),...,\hat{V}_{n}(\mathcal{S}(c,r_n))}$: Volumes for all particle neighborhoods embedded in $\mathcal{V}$}
 \State{$\mathcal{V}^{r_{\mathrm{max}}} = \{ \mathrm{x} \in \{ \mathcal{V} \land \mathcal{S}(c, r_{\mathrm{max}}) \} \}$}
 \State{$\mathcal{D} \gets \mathrm{distance\_transform(\mathcal{V}^{r_{\mathrm{max}}})}$}
 \ForAll{$r_i \in {r_1,...,r_n}$} 
  \State{$\mathrm{S} \gets \mathrm{gen\_binary\_mask(c, r_i, \mathcal{D})}$}
  \State{$\hat{V}_{i}(\mathcal{S}(c,r_i)) = \sum{\mathrm{S}}$}
 \EndFor
\end{algorithmic}\caption{\label{alg:dsa}3D array volume computation algorithm.}
\end{algorithm}

\begin{figure}
\begin{centering}
\includegraphics[width=1\textwidth]{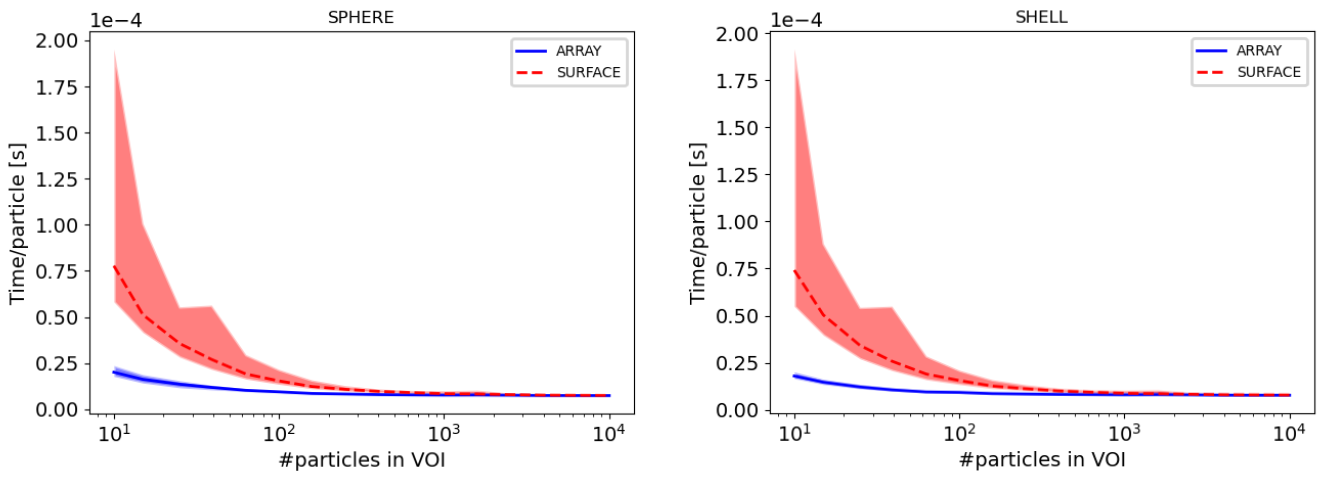}
\par\end{centering}
\caption{\label{fig:tm-num}Running time for evaluating the condition ${\ensuremath{\mathbf{x}\in\mathcal{V}}}$
(whether a particle belongs to VOI). Two different VOIs were used,
a sphere (left) and a thin spherical shell (right). The running times
per particle are shown for the surface and the 3D array representations
of VOI. The line center represents the median of $100$ simulations
and thickness the interval of confidence $\left[5,95\right]\%$.}
\end{figure}

\begin{figure}
\begin{centering}
\includegraphics[width=1\textwidth]{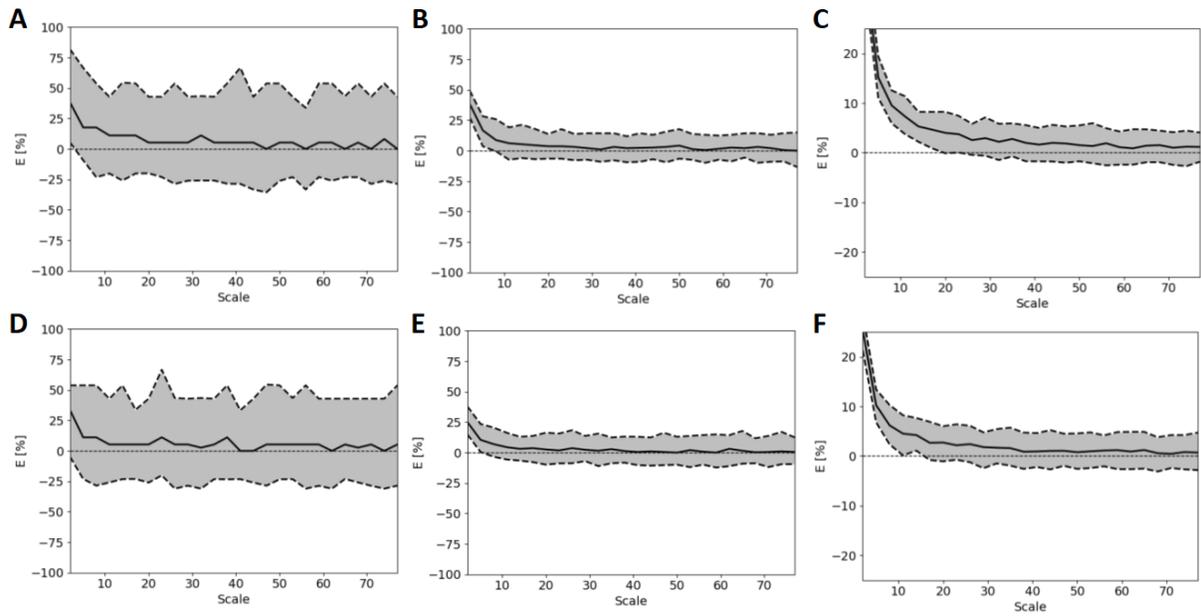}
\par\end{centering}
\caption{\label{fig:pr-MCS}Volume estimate precision for MCS algorithm. (A-C)
Half-spherical neighborhoods and (D-F) shell neighborhood, (A,D) $N=10$,
(B,E) $N=100$ and (C,F) $N=1000$, where $N$ is the number of points.
Solid lines are the median of 1000 executions and $IC=\left[5,95\right]\%$
in gray. $N_{rays}=10$.}
\end{figure}

\begin{figure}
\begin{centering}
\includegraphics[width=1\textwidth]{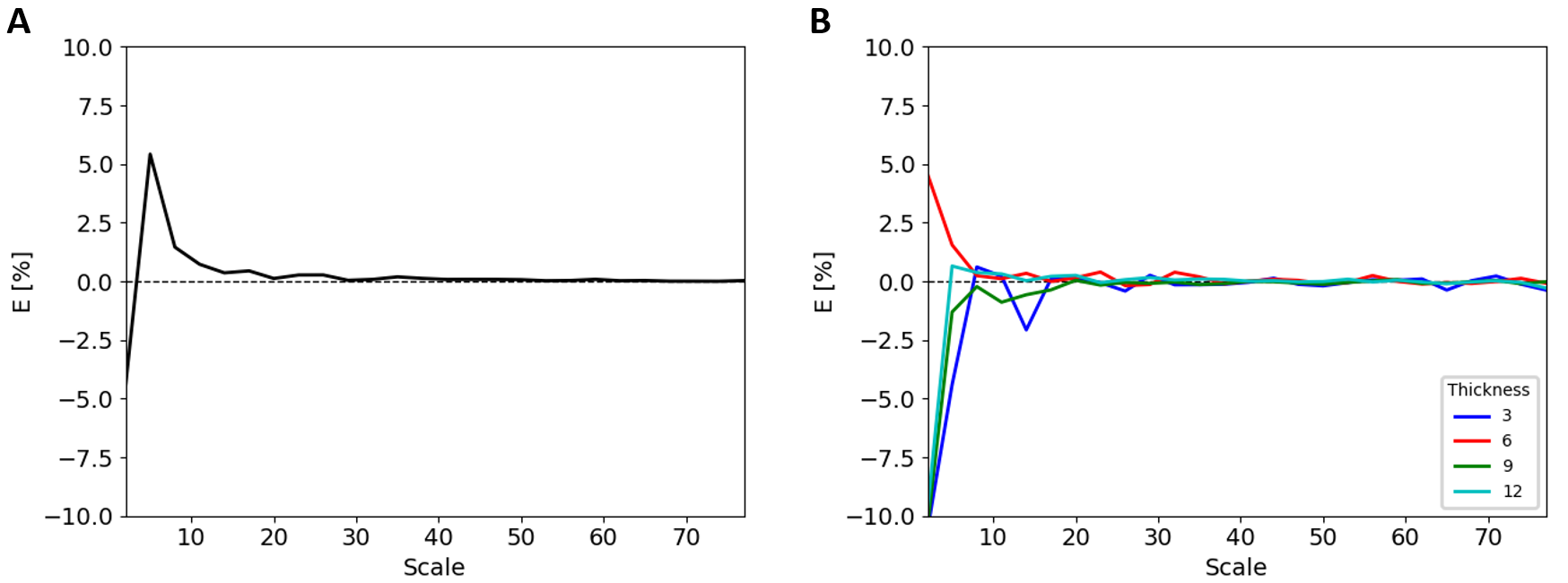}
\par\end{centering}
\caption{\label{fig:pr-DSA}Volume estimate precision for the 3D array algorithm.
A) Half-spherical neighborhoods B) Shell neighborhoods, the shell
thickness is indicated on the graph.}
\end{figure}

\begin{figure}
\begin{centering}
\includegraphics[width=1\textwidth]{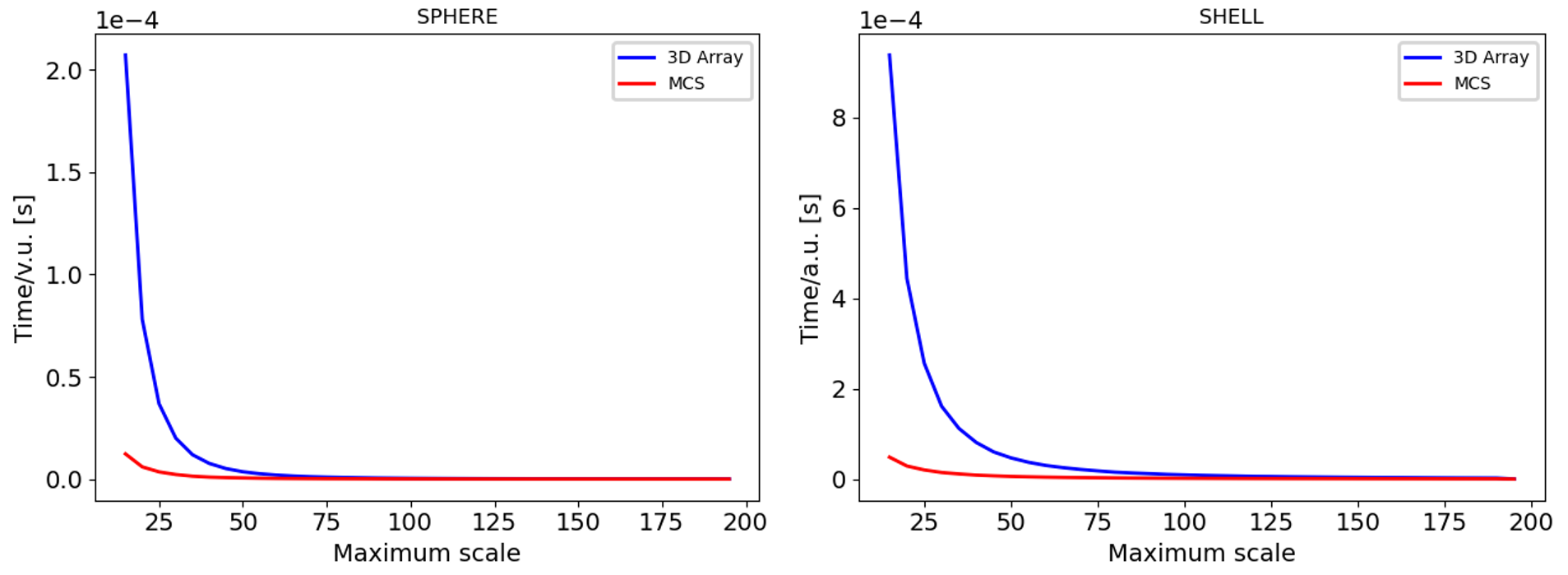}
\par\end{centering}
\caption{\label{fig:tm-MCS}Execution time comparison for volume determination
with MCS and 3D array algorithms. A) Running time per volume unit
for spherical neighborhoods B) Running time per area unit for shell
neighborhoods. In both cases points were randomly distributed. The
neighborhoods had different sizes, ranging from 15 to the maximum
size as indicated on X-axes.}
\end{figure}

\begin{figure}
\begin{centering}
\includegraphics[width=0.9\textwidth]{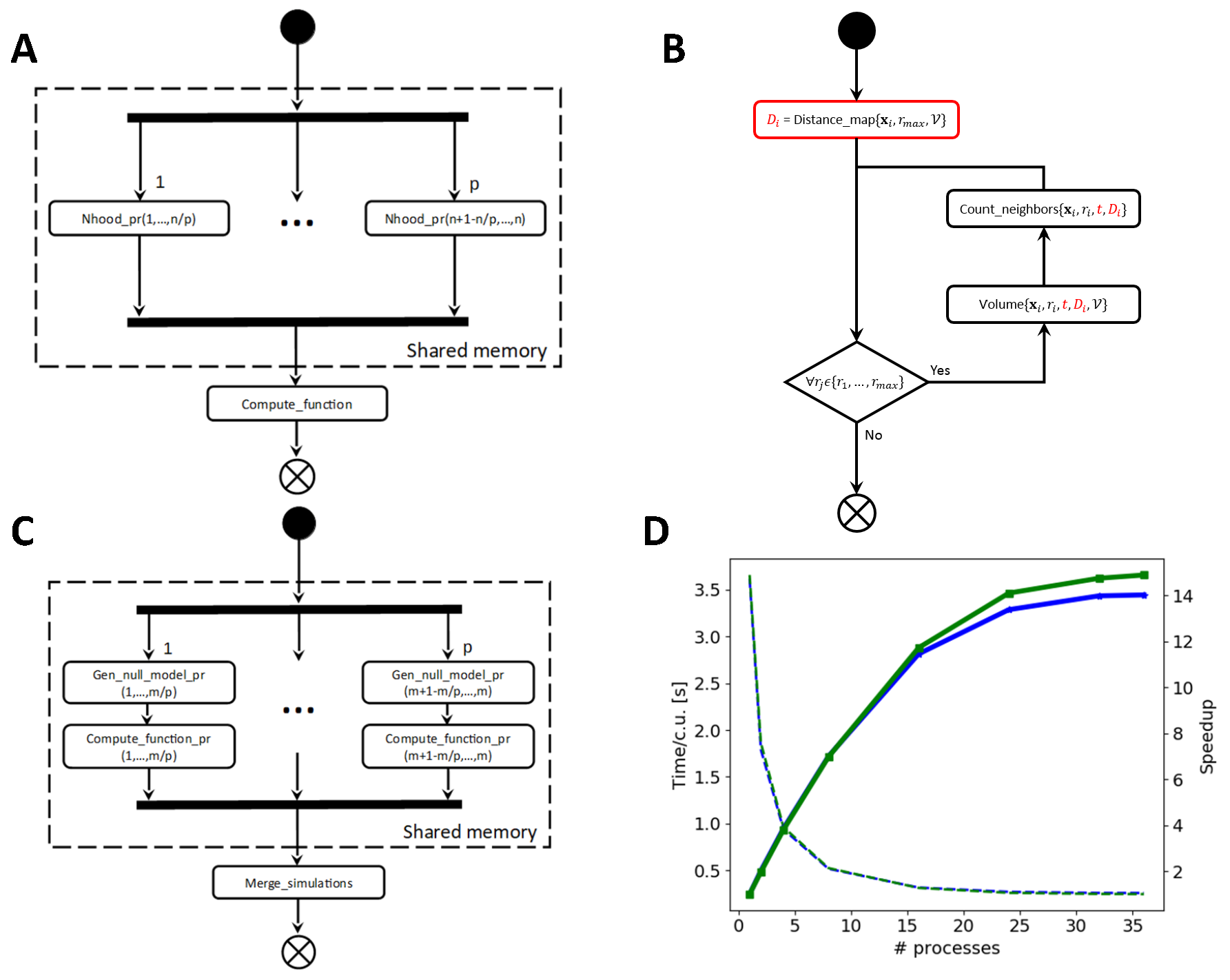}
\par\end{centering}
\caption{\label{fig:sch_nhood}Parallelization diagrams and execution speedup
for the second-order functions computations. A) Activity diagram for
processing real data using, $p\protect\leq n$ concurrent processes
($n$ is the number of particles). B) Diagram of tasks within a single
processing unit \textbf{\small{}(Nhood\_pr)} including the calculations
of the number of neighbors and the neighborhood l volumes for a range
of distances $\{r_{1},...,r_{max}\}$ (computing unit, c.u.), Operations
and parameters shown in red are only required for the 3D array but
not for MCS algorithm. C) Activity diagram for processing simulated
data using $p\protect\leq m$ processes concurrently ($m$ is the
total number of simulated tomograms). D) Execution time per c.u. (dashed
lines) and the speedup (solid lines) obtained with different amounts
of concurrent processes. Computations for experimental data (blue),
and simulations (green). Five synthetic tomograms with 500x500x100
pixels and 200 particles each were used for experimental data computation.
To ensure a fair comparison, the number of simulated tomograms on
each instance was the same as the number of concurrent processes.}
\end{figure}

\end{document}